\documentstyle{europhys}

\def\stars{\bigskip\centerline{***}\medskip}

\newif\ifboo \boofalse

\input psfig
\begin{document}

\euro{4x}{x}{x-x}{1999}
\Date{x xxxxxx 1999}
\shorttitle{P.~CH.~IVANOV {\it et al.}: SLEEP-WAKE DIFFERENCES IN SCALING 
BEHAVIOR}

\title{Sleep-Wake Differences in Scaling Behavior of the Human Heartbeat:
Analysis of Terrestrial and Long-Term Space Flight Data}

\author{Plamen~Ch.~Ivanov\inst{1,2}\footnote{Email:plamen@argento.bu.edu},
Armin Bunde\inst{3}, Lu\'{\i}s~A.~Nunes~Amaral\inst{1,2}, Shlomo
Havlin\inst{4}, Janice Fritsch-Yelle\inst{5}, Roman~M.~Baevsky\inst{6},
H.~Eugene~Stanley\inst{1}, and Ary~L.~Goldberger\inst{2}}

\institute{ 
	\inst{1} Center for Polymer Studies and Department of Physics,
  		Boston University, Boston, MA 02215, USA\\
	\inst{2} Cardiovascular Division, Harvard Medical School, 
		Beth Israel Deaconess Medical Center, Boston, MA 02215, USA\\
        \inst{3} Institute f\"ur Theoretische Physik III,
                Justus-Liebig-Universit\"at, Giessen, Germany\\           
	\inst{4} Gonda Goldschmid Center and Department of Physics, 
                Bar-Ilan University, Ramat Gan, Israel\\
	\inst{5} Life Sciences Research Laboratories,
                Lyndon B.\ Johnson Space Center, Houston, TX 77058, USA\\
        \inst{6} Institute of Biomedical Problems, Moscow, Russia
}

\rec{}{}

\pacs{
\Pacs{87}{.}{Biological and medical physics}
\Pacs{87}{19.Hh}{Cardiac dynamics}
\Pacs{87}{19.Jj}{Circadian rhythms}
\Pacs{87}{10.+e}{General theory and mathematical aspects}
}

\maketitle

\begin{abstract}
We compare scaling properties of the cardiac dynamics during sleep and
wake periods for healthy individuals, cosmonauts during orbital
flight, and subjects with severe heart disease. For all three groups,
we find a greater degree of anticorrelation in the heartbeat
fluctuations during sleep compared to wake periods.  The sleep-wake
difference in the scaling exponents for the three groups is comparable
to the difference between healthy and diseased individuals.  The
observed scaling differences are not accounted for simply by different
levels of activity, but appear related to intrinsic changes in the
neuroautonomic control of the heartbeat.
\end{abstract}

\begin{figure}
\centerline{\psfig{figure=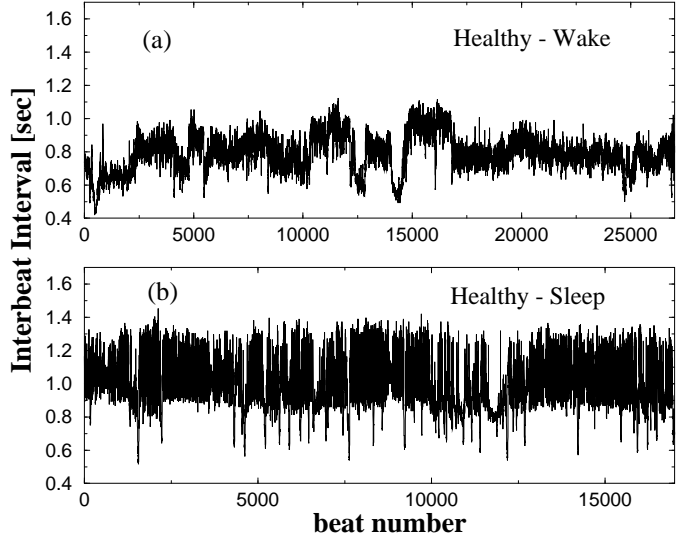,height=3.truein,angle=-90}}
\caption{ Consecutive heartbeat intervals are plotted vs beat number
for 6 hours recorded from the same healthy subject during: (a) wake
period: 12pm to 6pm and (b) sleep period: 12am to 6am.  (Note that
there are fewer interbeat intervals during sleep due to the larger
average of the interbeat intervals, i.e. slower heart rate.)}
\label{fig.data}
\end{figure}

The normal electrical activity of the heart is usually described as a
``regular sinus rhythm'' \cite{Bernard,Berne96}.  However, cardiac
interbeat intervals fluctuate in an irregular manner in healthy
subjects [fig.~\ref{fig.data}] --- even at rest or during sleep 
\cite{Ary96}.  The
complex behavior of the heartbeat manifests itself through the
nonstationarity and nonlinearity of interbeat interval sequences
\cite{Malik95}.  In recent years, the intriguing statistical properties
of interbeat interval sequences have attracted the attention
of researchers from different
fields~\cite{Mackey77,Wolf78,Kitney82,Glass91,Bassingthwaighte94,Kurths95,%
Sugihara96}.

Analysis of heartbeat fluctuations focused initially on short time
oscillations associated with breathing, blood pressure and
neuroautonomic control \cite{Kitney80,Akselrod81}.  Studies of longer
heartbeat records, however, revealed 1/f-like behavior
\cite{Kobayashi82,Saul87}.  Recent analysis of very long time series
(up to 24h: $n \approx 10^5$ beats) show that under healthy
conditions, interbeat interval increments exhibit power-law
anticorrelations \cite{Peng93}, follow a universal scaling form in
their distributions \cite{Ivanov96}, and are characterized by a broad 
multifractal spectrum \cite{Ivanov99}. 
These scaling features change with disease and advanced age
\cite{Lipsitz90}.  The emergence of scale-invariant properties in the
seemingly ``noisy'' heartbeat fluctuations is believed to be a result
of highly complex, nonlinear mechanisms of physiologic control
\cite{Shlesinger88}.

It is known that circadian rhythms are associated with periodic
changes in key physiological processes
\cite{Berne96,Malik95,Molgaard91,Ivanov98}.  
Here, we ask the question if there
are characteristic differences in the scaling behavior between sleep
and wake cardiac dynamics\footnote{Typically the differences in the
cardiac dynamics during sleep and wake phases are reflected in the
average (higher in sleep) and standard deviation (lower in sleep) of
the interbeat intervals \cite{Molgaard91}. Such differences can be
systematically observed in plots of the interbeat intervals recorded
from subjects during sleep and wake periods~[fig.~\ref{fig.data}].}.
We hypothesize that sleep and wake changes in cardiac control may
occur on all time scales and thus could lead to systematic changes in
the scaling properties of the heartbeat dynamics. Elucidating the
nature of these sleep-wake rhythms could lead to a better
understanding of the neuroautonomic mechanisms of cardiac regulation.

We analyze 30 datasets --- each with 24h of interbeat intervals ---
from 18 healthy subjects and 12 patients with congestive heart failure
\cite{MIT-BIH92}. We analyze the nocturnal and diurnal fractions of
the dataset of each subject which correspond to the 6h ($n \approx
22,000$ beats) from midnight to 6am and noon to 6pm.

We apply the detrended fluctuation analysis (DFA) method \cite{Peng94}
to quantify long-range correlations embedded in nonstationary
heartbeat time series. This method avoids spurious detection of
correlations that are artifacts of nonstationarity. Briefly, we first
integrate the interbeat-interval time series.  We then divide the time
series into boxes of length $n$ and perform, in each box, a
least-squares linear fit to the integrated signal. The linear fit
represents the local trend in each box.  Next, we calculate in each
box the root-mean-square deviations $F(n)$ of the integrated signal
from the local trend.  We repeat this procedure for different box sizes
(time scales) $n$. A power law relation between the average
fluctuation $F(n)$ and the number of beats $n$ in a box indicates the
presence of scaling; the correlations in the heartbeat fluctuations
can be characterized by the scaling exponent $\alpha$, defined as
$F(n) \sim n^{\alpha}$.  


\begin{figure}
\centerline{\psfig{figure=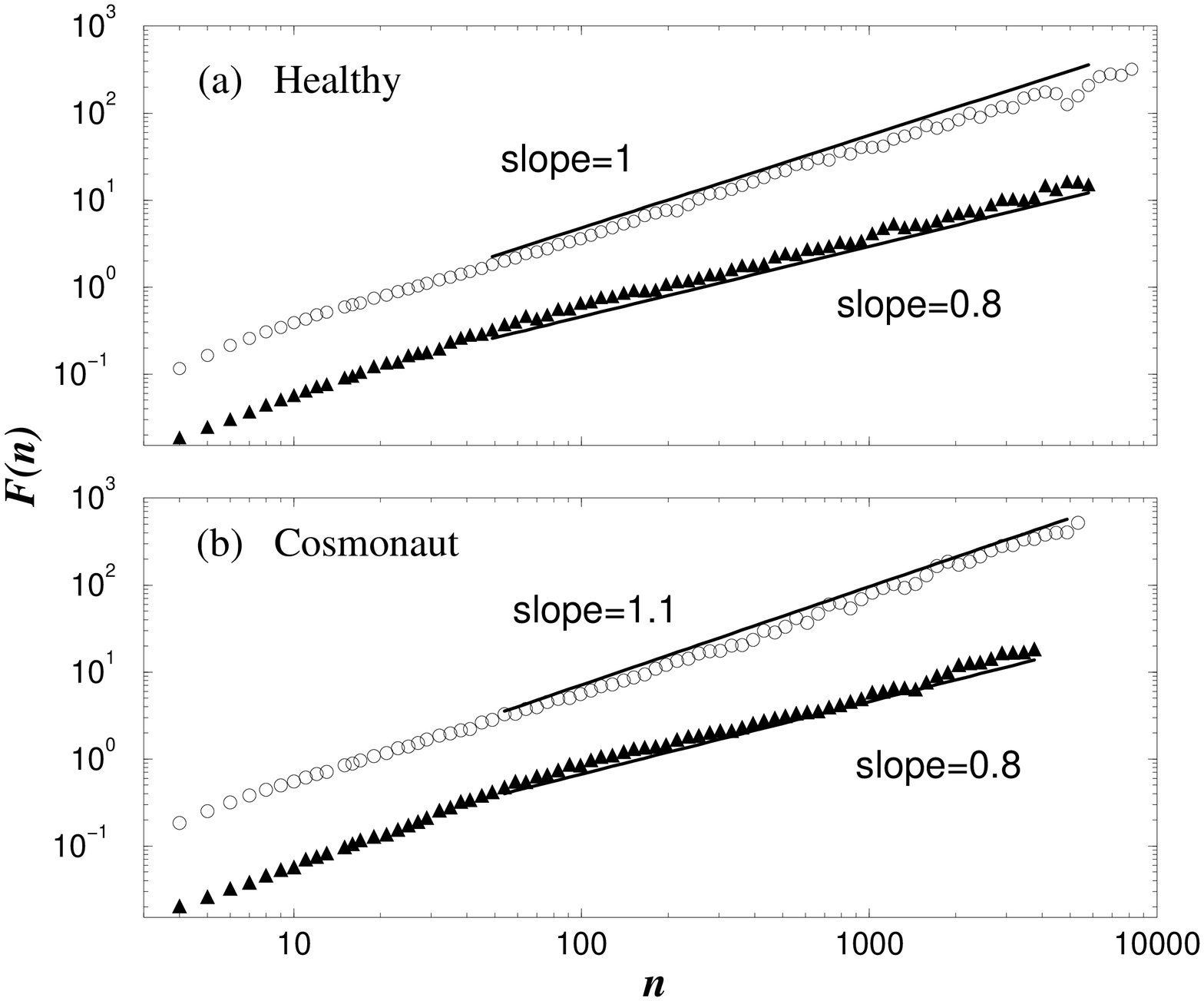,height=2.7truein,angle=-90}
\hspace*{-0.5cm}
\psfig{figure=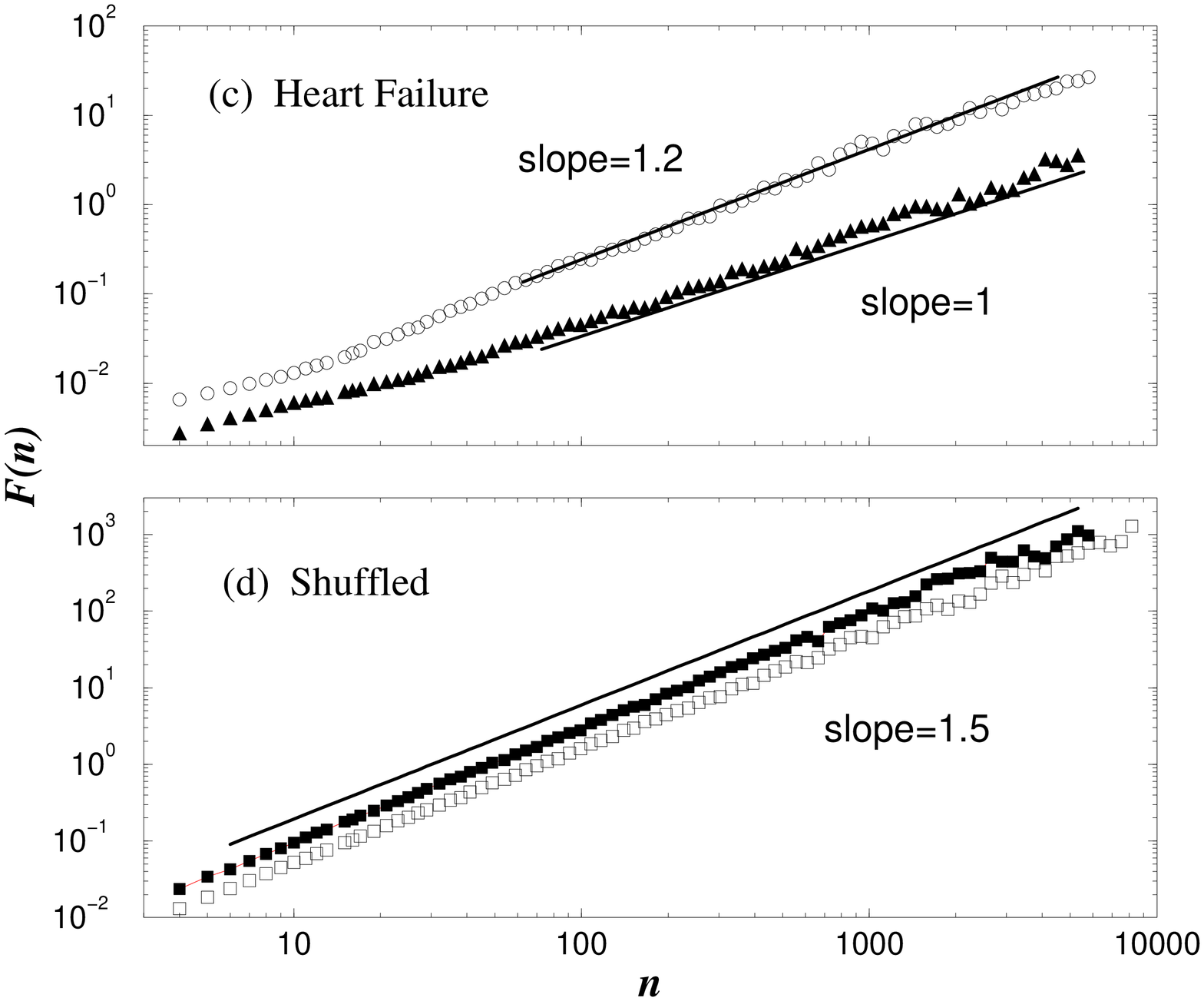,height=2.7truein,angle=-90}}
%
\caption{ Plots of $\protect\log F(n)$ vs. $\protect\log n$ for 6h
wake records (open circles) and sleep records (filled triangles) of (a) one
typical healthy subject; (b) one cosmonaut (during orbital flight);
and (c) one patient with congestive heart failure. Note the systematic
lower exponent for the sleep phase (filled triangles), indicating
stronger anticorrelations. (For some individuals we observe weak 
crossovers in the scaling of the fluctuation function $F(n)$
in the range $100<n<2000$.
However, there is no typical characteristic time scale $n$ at which
these crossovers occur and a crossover at given time scale $n$, 
for the wake data, does not appear to be associated
with a crossover, for the night data from the same individual. 
Such weak crossover events 
might be subject-specific and could be related to the particular
record of an individual, i.e. a repeat recording from the same
subject might not show such a crossover.)
(d) As a control, we reshuffle and
integrate the interbeat increments from the wake and sleep data of the
healthy subject presented in (a). We find a Brownian noise scaling
over all time scales for both wake and sleep phases with an exponent
$\alpha=1.5$, as one expects for random walk-like fluctuations. 
}
\label{fig.d_n.dfa}
\end{figure}

\begin{figure}
\centerline{\psfig{figure=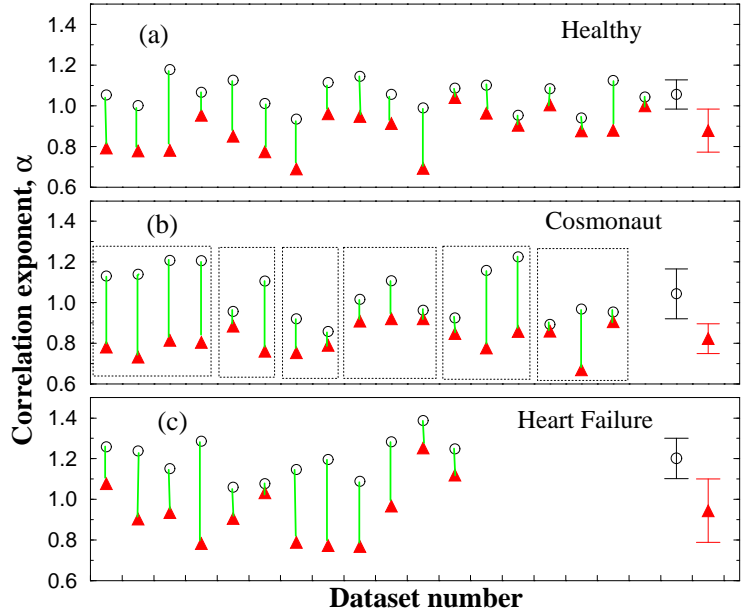,height=3.3truein,angle=-90}}
\caption{Values for the sleep (filled triangles) and wake activity
(open circles) exponents for all individual records of (a) the
healthy, (b) the cosmonaut, and (c) the heart failure groups. For the
healthy and heart failure groups each dataset corresponds to a
different individual.  Data from the 6 cosmonauts are grouped in 6
blocks, where each block contains data from the same individual,
recorded on different days during orbital flight and ordered from
early to late flight, ranging from the 3rd to the 158th day in orbit.
For all individuals in all groups, the day exponent exhibits
systematically a higher value than the sleep exponent. The sleep and
wake group averages and standard deviations are presented on the right
of each panel.}
\label{fig.groupexp}
\end{figure}

We find that at scales above $\approx 1$min ($n > 60$) the data during
wake hours display long-range correlations over two decades with
average exponents $\alpha_W \approx 1.05$ for the healthy group and
$\alpha_W \approx 1.2$ for the heart failure patients.  For the sleep
data we find a systematic crossover at scale $n~\approx~60$ beats
followed by a scaling regime extending over two decades characterized
by a smaller exponent: $\alpha_S \approx 0.85$ for the healthy group and
$\alpha_S \approx 0.95$ for the heart failure group
[fig.~\ref{fig.d_n.dfa}a,c].  Although the values of the sleep and
wake exponents vary from subject to subject, we find that for all
individuals studied, the heartbeat dynamics during sleep are
characterized by a smaller exponent [Table~\ref{t.statistics} and
fig.~\ref{fig.groupexp}].

\begin{table}
\caption{ Comparison of the statistics for the scaling exponents from the 
three groups in our database. Here, $N$ is the number of datasets in each 
group, $\alpha$ is the corresponding group average value and 
%
%
$\sigma$ is the standard deviation of the exponent values for each group.
%
%
The differences between the average sleep and wake phase exponents for
all three groups are statistically significant ($p<10^{-5}$ by the
Student's t-test).
}
\[
\begin{tabular}{clccccc}
& Group	              & $N$      & $\alpha$   & $\sigma$ \\
\hline
\hline
& Healthy Wake	      & $18$     & $1.05$       & $0.07$ \\
& Healthy Sleep       & $18$     & $0.85$       & $0.10$  \\
\hline
& Cosmonaut Wake      & $17$     & $1.04$       & $0.12$ \\
& Cosmonaut Sleep     & $17$     & $0.82$       & $0.07$  \\
\hline
& Heart Failure Wake  & $12$     & $1.20$        & $0.09$ \\
& Heart Failure Sleep & $12$     & $0.95$       & $0.15$ \\
%
\end{tabular}
\]
\label{t.statistics}
\end{table}

As a control, we also perform an identical analysis on two surrogate
data sets obtained by reshuffling and integrating the increments in
the interbeat intervals of the sleep and wake records from the same
healthy subject presented in fig.~\ref{fig.d_n.dfa}a. Both surrogate
sets display uncorrelated random walk fluctuations with a scaling
exponent of $1.5$ (Brownian noise) [fig.~\ref{fig.d_n.dfa}d]. The
value $1.5$ arises from the fact that we analyze the integral of the
signal, leading to an increase by 1 of the usual random-walk exponent
of $1/2$. A scaling exponent larger than $3/2$ would indicate
persistent correlated behavior, while exponents with values smaller
than $3/2$ characterize anticorrelations (a perfectly anticorrelated
signal would have an exponent close to zero).  Our results therefore
suggest that the interbeat fluctuations during sleep and wake phases
are long-range anticorrelated but with a significantly greater degree
of anticorrelation (smaller exponent) during sleep.

An important question is whether the observed scaling differences
between sleep and wake cardiac dynamics arise trivially from changes
in the environmental conditions (different daily activities are
reflected in the strong nonstationarity of the heartbeat time series).
Environmental ``noise'', however, can be treated as a ``trend'' and
distinguished from the more subtle fluctuations that may reveal
intrinsic correlation properties of the dynamics.  Alternatively, the
interbeat fluctuations may arise from nonlinear dynamical control of
the neuroautonomic system rather than being an epiphenomenon of
environmental stimuli, in which case only the fluctuations arising
from the intrinsic dynamics of the neuroautonomic system should show
long-range scaling behavior.

A possible explanation of the results from our analysis 
is that the observed sleep-wake scaling differences
are due to intrinsic changes in the cardiac control mechanisms for the
following reasons: (i) The DFA method removes the ``noise'' due to
activity by detrending the nonstationarities in the interbeat interval
signal related to polynomial trends 
and analyzing the fluctuations along the trends.  (ii)
Responses to external stimuli should give rise to a different type of
fluctuations having characteristic time scales, i.e. frequencies
related to the stimuli. However, fluctuations in both diurnal and
nocturnal cardiac dynamics exhibit scale-free behavior.  (iii) The
weaker anticorrelated behavior observed for all wake phase records
cannot be simply explained as a superposition of stronger
anticorrelated sleep dynamics and random noise of day activity.  Such
noise would dominate at large scales and should lead to a crossover
with an exponent of $1.5$.  However, such crossover behavior is not
observed in any of the wake phase datasets [fig.~\ref{fig.d_n.dfa}].
Rather, the wake dynamics are typically characterized by a stable
scaling regime up to $n=5\times10^3$ beats.

To test the robustness of our results, we analyze 17 datasets from 6
cosmonauts during long-term orbital flight on the Mir space station
\cite{NASA}.  Each dataset contains continuous periods of 6h data
under both sleep and wake conditions.  We find that for all cosmonauts
the heartbeat fluctuations exhibit an anticorrelated behavior with
average scaling exponents consistent with those found for the healthy
terrestrial group: $\alpha_W \approx 1.04$ for the wake phase and
$\alpha_S \approx 0.82$ for the sleep phase
[Table~\ref{t.statistics}].  This sleep-wake scaling difference is
observed not only for the group averaged exponents but for each
individual cosmonaut dataset [fig.~\ref{fig.d_n.dfa}b and
fig.~\ref{fig.groupexp}].  Moreover, the scaling differences are
persistent in time, since records of the same cosmonaut taken on
different days (ranging from the 3rd to the 158th day in orbit),
exhibit a higher degree of anticorrelation in sleep.

We find that even under the extreme conditions of zero gravity and
high stress activity, the sleep and wake scaling exponents for the the
cosmonauts are statistically consistent ($p=0.7$ by Student's t-test)
with those of the terrestrial healthy group\footnote{Those findings
are not inconsistent with the presence of other manifestations of
altered autonomic control during long-term spaceflight (T. Brown {\it
et al.\/}, preprint).}.  Thus, the larger values for the wake phase
scaling exponents cannot be a trivial artifact of
activity. Furthermore, the larger value of the average wake exponent
for the heart failure group compared to the other two groups
[Table~\ref{t.statistics}] cannot be attributed to external stimuli
either, since patients with severe cardiac disease are strongly
restricted in their physical activity.  Instead, our results suggest
that the observed scaling characteristics in the heartbeat
fluctuations during sleep and wake phases are related to intrinsic
mechanisms of neuroautonomic control.

The mechanism underlying heartbeat fluctuations may be related to
countervailing neuroautonomic inputs. Parasympathetic stimulation
decreases the heart rate, while sympathetic stimulation has the
opposite effect.  The nonlinear interaction between the two branches
of the nervous system is the postulated mechanism for the type of
complex heart rate variability recorded in healthy subjects
\cite{Berne96,Malik95}.  The fact that during sleep the scaling
exponents differ more from the value $\alpha =1.5$ (indicating
``stronger'' anticorrelations) may be interpreted as a result of
stronger neuroautonomic control.  Conversely, values of the scaling
exponents closer to $1.5$ (indicating ``weaker'' anticorrelations) for
both sleep and wake activity for the heart failure group are
consistent with previously reported pathologic changes in cardiac
dynamics \cite{Peng93}.  However, the average sleep-wake exponent
difference remains the same ($\approx 0.2$) for all three groups.  The
observed sleep-wake changes in the scaling characteristics may
indicate different regimes of intrinsic neuroautonomic regulation of
the cardiac dynamics, which may ``switch'' on and off associated with
circadian rhythms.

Surprisingly, we note that for the regime of large time scales ($n >
60$) the average sleep-wake scaling difference is comparable to the
scaling difference between health and disease;
cf.~Table~\ref{t.statistics} and\footnote{At small time scales ($n <
60$), we do not observe systematic sleep-wake differences. The scaling
exponents obtained from 24h records of healthy and heart failure
subjects in the asymptotic region of large time scales are in
agreement with the results for the healthy and heart failure groups
during the wake phase only. Since the weaker anticorrelations
associated with the wake phase are characterized by a larger exponent
while the stronger anticorrelated behavior during sleep has a smaller
exponent, at large scales the superposition of the two phases (in 24h
records) will exhibit behavior dominated by the larger exponent of the
wake phase.}.  We also note that the scaling exponents for the heart
failure group during sleep are close to the exponents observed for the
healthy group [Table~\ref{t.statistics}].  Since heart failure occurs
when the cardiac output is not adequate to meet the metabolic demands
of the body, one would anticipate that the manifestations of heart
failure would be most severe during physical stress when metabolic
demands are greatest, and least severe when metabolic demands are
minimal, i.e., during rest or sleep. The scaling results we obtain are
consistent with these physiological considerations: the heart failure
subjects should be closer to normal during minimal activity.  Of
related interest, recent studies indicate that sudden death in
individuals with underlying heart disease is most likely to occur in
the hours just after awakening \cite{stroke}. Our findings raise the
intriguing possibility that the transition between the sleep and wake
phases is a period of potentially increased neuroautonomic instability
because it requires a transition from strongly to weakly anticorrelated
regulation of the heart.

Finally, the finding of stronger heartbeat anticorrelations during
sleep is of interest from a physiological viewpoint, since it may
motivate new modeling approaches \cite{model,Ivanov-model98} and
supports a reassessment of the sleep phase as a surprisingly active
dynamical state. Perhaps the restorative function of sleep may relate
to an increased reflexive-type responsiveness of neuroautonomic
control, not just at one characteristic frequency, but over a broad
range of time scales.

\stars

We thank NIH/National Center for Research Resources (P41 RR13622), 
NASA, NSF, Israel-US Binational Science Foundation, 
The Mathers Charitable Foundation and FCT/Portugal for support.


\vskip-12pt
\begin{thebibliography}{99}

\bibitem{Bernard}
Bernard C., {\it Les Ph\'{e}nom\'{e}nes de la Vie} (Paris) 1878;
van der Pol B.\ and van der Mark J., {\it Phil. Mag} {\bf 6} (1928) 763;
Cannon W.\ B., {\it Physiol. Rev.} {\bf 9} (1929) 399.

\bibitem{Berne96} Berne R.\ M.\ and Levy M.\ N., {\it Cardiovascular
Physiology\/} 6th ed.  (C.V. Mosby, St. Louis) 1996.

\bibitem{Ary96} Goldberger A.\ L., {\it Lancet} {\bf 347} (1996) 1312.

\bibitem{Malik95} Malik M.\ and Camm A.\ J., eds. {\it Heart Rate
Variability\/} (Futura, Armonk NY) 1995.

\bibitem{Mackey77} Mackey M.\ and Glass L., {\it Science} {\bf 197} (1977) 287.

\bibitem{Wolf78} Wolf M.\ M.\, Varigos G. A., Hunt D., and Sloman J. G., 
{\it Med. J. Aust.} {\bf 2} (1978) 52.

\bibitem{Kitney82} Kitney R.\ I.\, Linkens D., Selman A. C., and 
McDonald A. A., {\it Automedica} {\bf 4} (1982) 141.

\bibitem{Glass91} Glass L., Hunter P.\ and McCulloch A., eds. {\it
Theory of Heart} (Springer Verlag, New York) 1991.

\bibitem{Bassingthwaighte94} Bassingthwaighte J.\ B., Liebovitch L.\
S.\ and West B.\ J., {\it Fractal Physiology} (Oxford Univ. Press, New
York) 1994.

\bibitem{Kurths95}
Kurths J., Voss A., Saparin P., Witf A., Kilner  H. J., and Wessel N.,
{\it Chaos} {\bf 5} (1995) 88.

\bibitem{Sugihara96}
Sugihara G.\, Allan W., Sobel D., and Allan K. D.,
{\it Proc. Natl. Acad. Sci. USA} {\bf 93} (1996) 2608.

\bibitem{Kitney80}
Kitney R.\ I.\ and Rompelman O., {\it The Study of Heart-Rate Variability}
(Oxford Univ. Press, London) 1980.

\bibitem{Akselrod81}
Akselrod S.\, Gordon D., Ubel F. A., Shannon D. C., Barger A. C., and 
Cohen R. J., {\it Science} {\bf 213} (1981) 220.

\bibitem{Kobayashi82}
Kobayashi M.\ and Musha T., {\it IEEE Trans. Biomed. Eng.} 
{\bf 29} (1982) 456. 


\bibitem{Saul87}
Saul J.\ P.\, Albrecht P., Berger D., and Cohen R. J., 
{\it Computers in Cardiology} (IEEE Computer Society Press, Washington
DC), (1987) 419.


\bibitem{Peng93} 
Peng C.-K. {\it et al.\/}, 
{\it Phys. Rev. Lett.} {\bf 70} (1993) 1343; 
Peng C.-K., Havlin S., Stanley H.\ E.\ and Goldberger A.\
L., {\it Chaos} {\bf 5} (1995) 82; Turcott R.\ G.\ and Teich M.\ C., 
{\it Ann. of Biomed. Eng.} {\bf 24} (1996) 269.


\bibitem{Ivanov96} 
Ivanov P.\ Ch., Rosenblum M.G., Peng C.-K., Mietus J., 
Havlin S., Stanley H. E., and Goldberger~A.~L., 
{\it Nature} {\bf 383} (1996) 323.

\bibitem{Ivanov99} 
Ivanov P.\ Ch., Amaral L. A. N., Goldberger A. L., Havlin S., 
Rosenblum M. G., Struzik Z. R., Stanley H. E., 
{\it Nature} {\bf 399} (1999) 461.




\bibitem{Lipsitz90}
Lipsitz L.\ A.\, Mietus J., Moody G. B., and Goldberger A. L.,
{\it Circulation} {\bf 81} (1990) 1803; Kaplan D.\ T.\ {\it et al.\/},
{\it Biophys. J.} {\bf 59} (1991) 945; Iyengar N.\ {\it et al.\/},
{\it Am. J. Physiol.} {\bf 271} (1996) R1078.

\bibitem{Shlesinger88} Shlesinger M.\ F.\ and West B.\ J., {\it Random
Fluctuations and Pattern Growth: Experiments and Theory}, Stanley H. E.
and Ostrowsky N., eds.
(Proceedings 1988 Carg\`ese NATO ASI Series E: Applied Sciences, Vol 157). 
(Kluwer Academic Publishers, Boston) 1988.


\bibitem{Molgaard91}
Moelgaard H.\, Soerensen K. E., and Bjerregaard P.
{\it Am. J. Cardiol.} {\bf 68} (1991) 777.

\bibitem{Ivanov98}
Ivanov P.\ Ch. {\it et al.\/}, 
{\it Physica A} {\bf 249} (1998) 587.

\bibitem{MIT-BIH92} {\it Heart Failure Database\/} (Beth Israel
Deaconess Medical Center, Boston, MA).  The database now includes 18
healthy subjects (13 female and 5 male, with ages between 20 and 50,
average 34.3 years), and 12 congestive heart failure subjects (3
female and 9 male, with ages between 22 and 71, average 60.8 year) in
sinus rhythm.

\bibitem{Peng94}
Peng~C.-K.\,~Buldyrev~S.~V.,~Havlin~S.,~Simons~M.,~Stanley~H.~E., and
Goldberger A. L., {\it Phys. Rev. E}~{\bf 49}~(1994)~1685; Peng C.-K.,
Hausdorff J. M., Goldberger A. L., {\it Nonlinear Dynamics,
Self-Organization, and Biomedicine}, Walleczek J, ed. 
(Cambridge University Press,Cambridge,1999); see the online version at
http://reylab.bidmc.harvard.edu/DynaDx/index.html.


\bibitem{NASA}
Goldberger A.\ L.\ {\it et al.\/},
{\it Am. Heart J.} {\bf 128} (1994) 202.

\bibitem{stroke} Peters R.\ W.\ {\it et al.\/}
{\it J. Am. Coll. Cardiol.} {\bf 23} (1994) 283;
Behrens S.\ {\it et al.\/} {\it Am. J. Cardiol.} {\bf 80} (1997) 45.

\bibitem{model} Key statistical characteristics of the healthy cardiac
dynamics can be successfully reproduced by a stochastic nonlinear
feedback mechanism \cite{Ivanov-model98}. The present observation of
sleep-wake scaling differences poses a new challenge to such modeling
approaches, which could require considering reciprocity in the
activity of the sympathetic and parasympathetic branches of the
autonomic nervous system during sleep and wake phases, as well as
different correlation times of the sympathetic and parasympathetic
impulses.

\bibitem{Ivanov-model98} 
Ivanov P.\ Ch., Amaral L.\ A.\ N., Goldberger A.\ L.\ and Stanley H.\ E.,
{\it Europhys. Lett.} {\bf 43} (1998) 363.

\end{thebibliography}
\end{document}